\documentclass[journal]{IEEEtran}
\usepackage{amsmath,amsfonts}
\usepackage{algorithmic}
\usepackage{algorithm}
\usepackage{array}
\usepackage[caption=false,font=normalsize,labelfont=sf,textfont=sf]{subfig}
\usepackage{textcomp}
\usepackage{stfloats}
\usepackage{url}
\usepackage{verbatim}
\usepackage{graphicx}
\usepackage{cite}
\usepackage{xfrac}
\usepackage{siunitx}
\usepackage{tabularx}
\usepackage{hyperref}
\usepackage{multirow}
\usepackage{graphicx}
\usepackage{pifont}
\begin{document}

\title{Memristor-Based Selective Convolutional Circuit for High-Density Salt-and-Pepper Noise Removal}

\author{
  Binghui Ding, Ling Chen, Chuandong Li, \IEEEmembership{Senior Member, IEEE}, \\
  Tingwen Huang, \IEEEmembership{Fellow, IEEE}, and Sushmita Mitra, \IEEEmembership{Fellow, IEEE}
  \thanks{This work is supported by the National Key Research and Development Project (2018AAA0100101), and the National Natural Science Foundation of China (62373310). It is also supported by the Chongqing Municipal Education Commission (Grant no. KJQN202300207), and Chongqing Higher Education Teaching Reform (Grant no. 233094).}
  \thanks{Binghui~Ding, Ling~Chen and Chuandong~Li are with the College of Electronic and Information Engineering, Southwest University, Chongqing 400715, China (e-mail:binghui.ding99@gmail.com; 2006chenling2006@163.com; cdli@swu.edu.cn).}
  \thanks{Tingwen Huang is with Science Program, Texas A\&M University at Qatar, Doha 23874, Qatar (e-mail: tingwen.huang@qatar.tamu.edu).}
  \thanks{Sushmita Mitra is with Machine Intelligence Unit, Indian Statistical Institute, Kolkata 700108, India (e-mail: sushmita@isical.ac.in).}
}

\maketitle

\begin{abstract}
  In this article, we propose a memristor-based selective convolutional (MSC) circuit for salt-and-pepper (SAP) noise removal. We implement its algorithm using memristors in analog circuits. In experiments, we build the MSC model and benchmark it against a ternary selective convolutional (TSC) model. Results show that the MSC model effectively restores images corrupted by SAP noise, achieving similar performance to the TSC model in both quantitative measures and visual quality at noise densities of up to 50\%. Note that at high noise densities, the performance of the MSC model even surpasses the theoretical benchmark of its corresponding TSC model. In addition, we propose an enhanced MSC (MSCE) model based on MSC, which reduces power consumption by 57.6\% compared with the MSC model while improving performance.
\end{abstract}


\section{PREFACE}
\label{s1}
\IEEEPARstart{I}{mage} denoising aims to eliminate the noise interference encountered in the process of image acquisition, compression, and transmission while preserving the edges, textures, and details of the image\cite{xu2017fast}. The primary objective is to obtain high-quality denoised images to provide robust support for image processing and computer vision tasks \cite{ghimpecteanu2015decomposition}. Impulse noise, as a prevalent interference in images, randomly changes the values of a part of pixels. Impulse noise exists in two forms: fixed-valued noise and random-valued noise, with fixed-valued noise also known as salt-and-pepper (SAP) noise. In the presence of SAP noise, noisy pixels have a probability (denoted by \(D_s\)) to take the maximum value (255 in 8-bit images) or a probability (denoted by \(D_p\)) to take the minimum value (0). These two cases are respectively referred to as salt noise (white dots) and pepper noise (black dots). Typically, \(D_s\) and \(D_p\) are considered equal, while \(D = D_s + D_p\) represents the noise density.

Numerous methods for SAP noise removal have been proposed, including median filters and their variants \cite{li2013computation,khateb20150,zhang2018modified,hwang1995adaptive,meher2014improved,erkan2018different,monajati2015approximate}, mean filters and their variants \cite{memics2021different,zhang2014new,thanh2020adaptive,erkan2020improved}, combinations of median and mean filters \cite{erkan2018different,enginouglu2020adaptive}, and interpolation filters \cite{varatharajan2018adaptive}. Among them, significant attention has been garnered by median-based filters due to the pixel corruption caused by SAP noise, which replaces pixel values with either the maximum or minimum value. Khateb et al. \cite{khateb2016low} proposed a low-voltage low-power voltage-mode analog median filter based on winner-take-all and loser-take-all circuits. In addition, The median filter based on sorting networks\cite{monajati2019modified}, compared to the imprecise algorithm in \cite{monajati2015approximate}, enabling more effective removal of SAP noise while reducing hardware resource utilization \cite{vasicek2016evolutionary}. The architecture demonstrates good performance when evaluated against the peak signal-to-noise ratio (PSNR) and the structural similarity index measure (SSIM) criterion at 10\% noise density (PSNR=32.52dB, SSIM=0.98). For images corrupted by 15\% and 30\% SAP noise, median filtering employed 26 maximum/minimum selectors\cite{yildirim2021analog} based on an enhanced dual-rank sorting network, resulting in PSNR/SSIM values of 47.78dB/0.976 and 45.93dB/0.977, respectively.

Convolutional neural networks (CNNs) has good effect on the Gaussian image noise removal\cite{zhang2018residual,zhang2021plug,valsesia2020deep,zhang2018ffdnet,zhang2017beyond}. However, the performance is suboptimal, particularly for SAP noise, which is irrelevant to the original pixels, especially at high noise densities \cite{zhang2015salt,xing2019deep}. To overcome these drawbacks, the multistage selective convolution filter (MSCF) \cite{rafiee2023very} and the selective convolutional neural network (SeConvNet) \cite{rafiee2023deep} have been proposed. SeConvNet achieved high PSNR/SSIM values of 44.30dB/0.995 on the 20 traditional test images dataset at 10\% noise density. Even at 95\% noise density, SeConvNet maintained good performance with PSNR/SSIM values of 25.88dB/0.833.

Our interest has been piqued by the potential of utilizing memristive circuits to implement SeConvNet, aiming to address the research deficiencies in high-density SAP noise removal circuits. Despite the existing research on memristor-based SAP noise removal circuits, such as the denoising autoencoder network\cite{suresh2019realizing} utilizing Cu:ZnO memristors, the noise removal memristive circuit\cite{zarandi2020memristor} employing a strategy based on distance constraints between adjacent pixels for calibration, the circuit utilizing spin memristor cross arrays\cite{zhu2020convolution} for restoring SAP noise images, and the self-renewing mask circuit (SRMC)\cite{shang2018srmc} capable of reducing memory/communication burden during noise removal. However, these circuits are usually only effective at low noise density (approximately 10\%).

In light of the current lack of an efficient memristive neuron circuit for high-density SAP noise removal, we propose to construct a memristor-based SeConvNet circuit. Memristors offer remarkable advantages in hardware implementations of neural networks, particularly in neuromorphic computing, matrix operations, and power efficiency \cite{duan2024memristor}. They can be employed in crossbar arrays for synaptic weight storage \cite{kataeva2015efficient,krestinskaya2019neuromemristive}. Therefore, the integration of memristor and deep learning has considerable promise\cite{alibart2013pattern,hasan2014enabling,soudry2015memristor,zhang2023edge}.

The existing SAP noise removal circuits present two issues.

\begin{enumerate}
  \item CNNs have demonstrated notable performance enhancements compared to classical algorithms. However, their tendency to globally modify image pixels may result in suboptimal restoration of non-additive noise with pixel values unrelated to the original ones, such as SAP noise, potentially leading to performance degradation. In addition, the residual learning strategies in models may introduce undesirable visual artifacts \cite{radlak2020deep}.

  \item As memristor technology advances, an increasing number of novel networks or algorithms are being employed in memristor-based circuits for SAP noise removal. However, existing memristor circuits typically demonstrate effectiveness only in restoring SAP noise images at low noise density (approximately 10\%).
\end{enumerate}

The SeConvNet model, designed for removing SAP noise in images, is characterized by its ability to effectively restore images corrupted by SAP noise, even at high noise densities. An efficient SeConvNet model based on memristors is eagerly awaited in research circles. In this paper, a memristor-based SeConvNet circuit for SAP noise removal is proposed, with emphasis placed on the circuit implementation method of its core algorithms.

The primary contributions of our study are as follows.

\begin{enumerate}
  \item We employed memristors to implement the algorithms of SeConvNet within analog circuits.

  \item We established simulations of the circuit in PSpice. Testing was conducted on images at different SAP noise densities to validate our work. The results indicate that the circuit exhibits robust performance even at 50\% noise density.
  
  \item We improved the algorithm of SeConvNet and proposed an enhanced model, which reduces power consumption by 57.6\% while improving performance.
\end{enumerate}

We compared the output of the circuit model with theoretical values. The results indicate that the discrepancy between the two is negligible when the noise density is less than or equal to 50\%. However, when the noise density is greater than or equal to 60\%, the performance of the circuit model surpasses the theoretical values. We conducted an analysis and discussion on this matter.

The paper is organized as follows. Section \ref{s1} reviews various types of SAP noise removal circuits, identifies current research deficiencies, and outlines the strategies and experimental designs. Section \ref{s2} introduces the principles of SeConvNet and fundamental knowledge in the relevant field. Section \ref{s3} presents memristor-based SAP noise removal circuits in detail. Section \ref{s4} discusses the experimental methodology and results. Section \ref{s5} concludes our work and explores future research directions.

\section{PRELIMINARIES}
\label{s2}
The SeConvNet \cite{rafiee2023deep} consists of selective convolutional (SeConv) layers and traditional convolutional layers. The SeConv layer has a single channel with kernel size of \(s\times s\), where s=3, 5, 7, 9, 11, 13, 15. The traditional convolutional layers have 64 channels with kernel size of \(3\times 3\). In addition, the output channels of the last convolutional layer match the channels of the input image to ensure the accuracy of image reconstruction. 

Before being input into the SeConv model, the tensor A normalized by noisy image needs preprocessing to convert all potential noisy pixel values in tensor \(A\) to 0, resulting in \(\widetilde{A}\), i.e.,

\begin{equation}
  \label{eq1}
  \left[\widetilde{A}\right]_{i,j,k} =
  \begin{cases}
    0 & \text{if } \left[A\right]_{i,j,k}=1 \\
    \left[A\right]_{i,j,k} & \text{otherwise},
  \end{cases}
\end{equation}

\noindent where \(\left[A\right]_{i,j,k}\) is the element value of tensor \(A\) at coordinates \((i,j,k)\). This process turns all noisy pixel values become 0, while non-zero values remain unchanged as uncorrupted pixels. Setting the non-zero pixel values in \(\widetilde{A}\) to 1, the non-noisy pixels map of tensor \(\widetilde{A}\) is obtained as

\begin{equation}
  \label{eq2}
  \left[{\widetilde{M}}_A\right]_{i,j,k} =
  \begin{cases}
    1 & \text{if } \left[A\right]_{i,j,k}\neq0 \\
    0 & \text{otherwise}.
  \end{cases}
\end{equation}

Tensor \(A\) is preprocessed to derive \(\widetilde{A}\) as the input for the SeConv model. \(\widetilde{A}\) is passed through a gain limiter, resulting in \(\widetilde{M}_A\). Details of this circuitry are omitted. Subsequently, the normalized weighted average of non-noisy pixels around an \(s \times s\) window is calculated by convolving between the input tensor \(\widetilde{A}\) and the kernel \cite{thanh2019iterative}, i.e.,

\begin{equation}
  \label{eq3}
  \left[N\right]_{i,j,k} =
  \begin{cases}
    \frac{\left[{\widetilde{A}}_{conv}\right]_{i,j,k}}{\left[{\widetilde{M}}_{Aconv}\right]_{i,j,k}} & \text{if } \left[{\widetilde{M}}_{Aconv}\right]_{i,j,k}\neq0 \\
    0 & \text{otherwise},
  \end{cases}
\end{equation}

\noindent where \({\widetilde{A}}_{conv}\) and \({\widetilde{M}}_{Aconv}\) represent the results of \(\widetilde{A}\) and \({\widetilde{M}}_A\) after traditional convolution.

Valuable noise pixels can be selectively restored by setting reliability tensors,

\begin{equation}
  \label{eq4}
  \left[F_M\right]_{i,j,k} = 
  \begin{cases}
    1 & \text{if } \left[{\widetilde{F}}_{Mconv}\right]_{i,j,k}\geq\eta \\
    0 & \text{otherwise},
  \end{cases}
\end{equation}

\noindent where \(\eta = s - 2\). The convolution kernel with uniform weight of 1 is defined as the fixed convolution kernel. Thus, \({\widetilde{F}}_{Mconv}\) represents the result of \({\widetilde{M}}_A\) after fixed convolution. Inverting 0s and 1s in \({\widetilde{M}}_A\) yields the noisy pixels map \(M_A\) of tensor \(\widetilde{A}\), i.e.,

\begin{equation}
  \label{eq5}
  \left[M_A\right]_{i,j,k}=1-\left[{\widetilde{M}}_A\right]_{i,j,k}.
\end{equation}

The product of \(N\), \(M_A\), and \(F_M\) is the restored pixels corresponding to the noisy pixels. The restored tensor of the SeConv model is obtained as

\begin{equation}
  \label{eq6}
  \hat{A}=\widetilde{A}+N\odot M_A\odot F_M,
\end{equation}

\noindent where \(\odot\) denotes the element-wise product. Restore image is obtained by multiplying \(\hat{A}\) by 255.

The SeConv layer assigns a single channel to each feature, with the first layer supporting three channels in the case of RGB image input. This inherent design provides natural advantages for hardware circuitry. In addition, memristors allow for large-scale parallel storage and computation at low power\cite{aguirre2024hardware}. Hence, the memristor-based SeConvNet circuit shows potential for efficient image processing with low power consumption.

\begin{figure}[!t]
  \centering
  \includegraphics[width=3.4in]{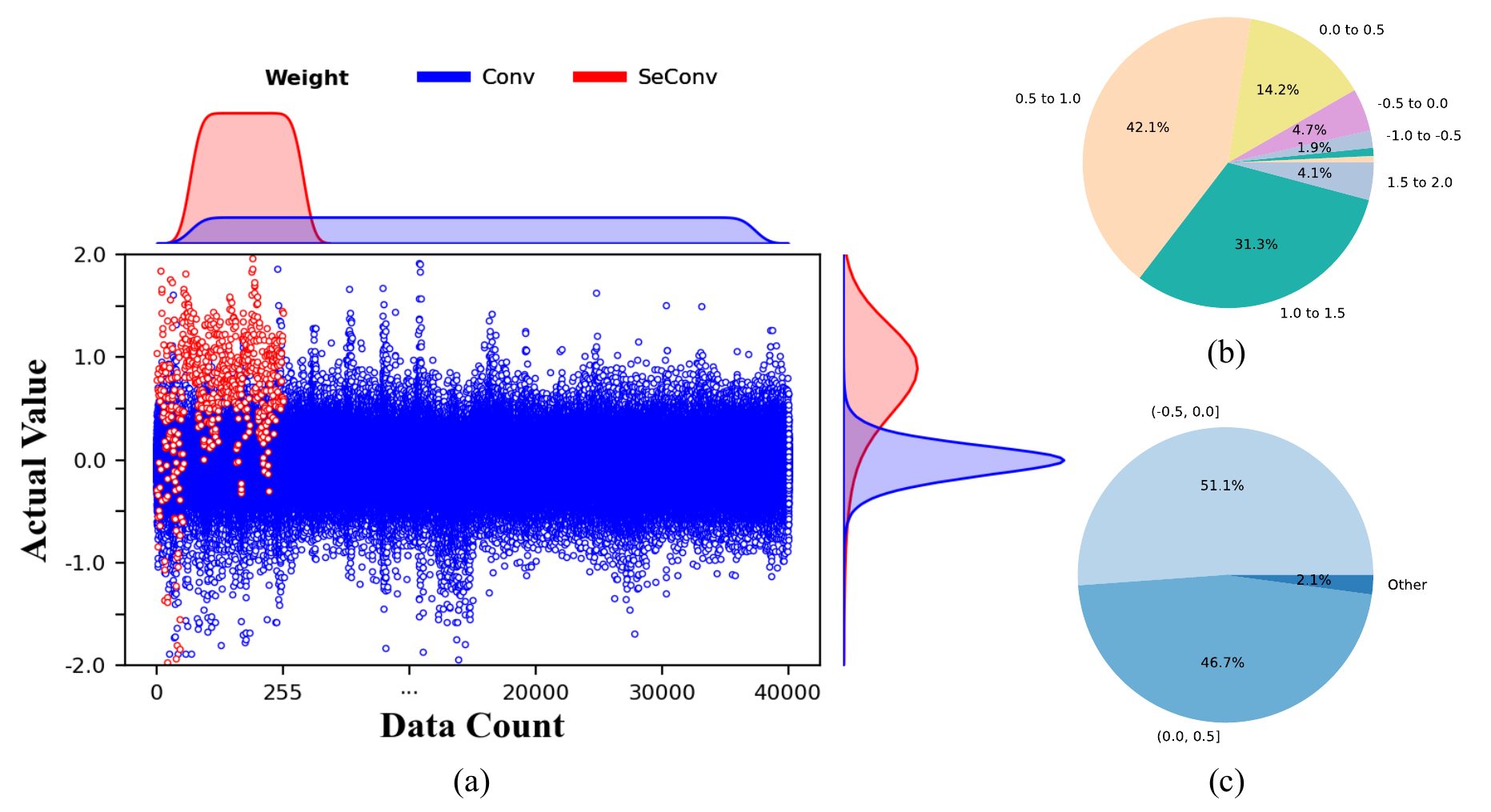}
  \caption{The weight distributions of traditional convolutional layers (blue scatter points) and SeConv layers (red scatter points) in SeConvNet are shown in (a), while the distribution of weight counts and values is displayed through kernel density estimation plots located above and to the right, respectively. Normal distribution properties are observed in both traditional convolutional layers and SeConv layers. The proportion of weight ranges for SeConv layers and traditional convolutional layers is depicted in (b) and (c) respectively.}
  \label{fig_1}
\end{figure}

For achieving a stable and controllable implementation of SeConvNet in memristive circuits, despite the inherent presence of multiple intermediate states in memristors, we opt to utilize only two distinct memristor states: the high resistance state (HRS) and low resistance state (LRS). Consequently, the quantization of full precision weights becomes necessary. As depicted in Fig. \ref{fig_1}, the weights of SeConv layers and traditional convolutional layers can be approximated as normal distributions. Thus, we adopt a ternary quantization strategy similar to Liu et al. \cite{li2016ternary}, where the ternary weights \(\widetilde{W}\) are computed by

\begin{equation}
  \label{eq7}
  \begin{aligned}
    \widetilde{W}_i = f(W_i | \theta) =
    \begin{cases}
      +1 & \text{if } W_i > \theta   \\
      0  & \text{if } |W_i| < \theta \\
      -1 & \text{if } W_i < -\theta,
    \end{cases}
  \end{aligned}
\end{equation}

\noindent where the positive threshold parameter \(\theta\) is as follows,

\begin{equation}
  \label{eq8}
  \theta=0.75\ast E\left(\left|W\right|\right)=\frac{0.75}{n}\sum_{i=1}^{n}\left|W_i\right|.
\end{equation}

The memristor model with threshold\cite{pershin2012spice} employed in this study can be expressed as

\begin{equation}
  \label{eq9}
  i_{\left(t\right)}=R_{\left(t\right)}^{-1}v_{\left(t\right)}
\end{equation}

\begin{equation}
  \label{eq10}
  \begin{aligned}
  \frac{dR_{\left(t\right)}}{dt} = f_{th}\left(v_{\left(t\right)}\right)\left[u\left(v\left(t\right)\right)u\left(R_{OFF}-R_{\left(t\right)}\right) \right.& \\
  \left. +u\left(-v_{\left(t\right)}\right)u\left(R_{\left(t\right)}-R_{ON}\right)\right]&,
  \end{aligned}
\end{equation}

\noindent with

\begin{equation}
  \label{eq11}
  f_{th}\left(v_{\left(t\right)}\right)=\beta v_{\left(t\right)}+0.5\left(\alpha-\beta\right)\left[\left|v_{\left(t\right)}+V_{th}\right|-\left|v_{\left(t\right)}-V_{th}\right|\right],
\end{equation}

\noindent where \(R_{OFF}\) and \(R_{ON}\) are the resistance of the memristor in HRS and LRS respectively. \(V_{th}\) is the threshold voltage, \(u(\cdot)\) is the step functions, and the \(\alpha\) and \(\beta\) is the change rate of memristance. Note that the device changes only if \(\left|v_{\left(t\right)}\right|>V_{th}\) at \(\alpha\)=0.

The trinary weight \({\widetilde{W}}_i\) can be mapped by the conductance \(G_i\) of memristors in the circuit. When \(G_{ON}\gg G_{OFF}\), \(G_i\) can be represented by \(G_i^+\) and \(G_i^-\) as

\begin{equation}
  \label{eq12}
  G_i = G_i^+ - G_i^- =
  \begin{cases}
    G_{ON} - G_{OFF}  & \text{if } W_i = 1   \\
    G_{OFF} - G_{OFF} & \text{if } W_i = 0   \\
    G_{OFF} - G_{ON}  & \text{if } W_i = -1,
  \end{cases}
\end{equation}

\noindent where \(G_{ON} = 1/R_{ON}\), and \(G_{OFF} = 1/R_{OFF}\). In this study, \(R_{ON}\) is \(10k\Omega\), \(R_{OFF}\) is \(1M\Omega\), \(V_{th}\) is 1.5V, \(\alpha\) is 0, and \(\beta\) is \(10^{7}\).

\section{MEMRISTOR-BASED SECONV CIRCUITS FOR SALT-AND-PEPPER NOISE REMOVAL}
\label{s3}
\subsection{Overview}
Given the extensive research on circuit designs for traditional convolutional and batch normalization (BN) layers \cite{ran2020shufflenetv2,ran2020memristor,chen2021highly,wen2018memristive,wen2019memristor}, we focus on the design of the SeConv model.

\begin{figure*}[!t]
  \centering
  \includegraphics[width=7.0in]{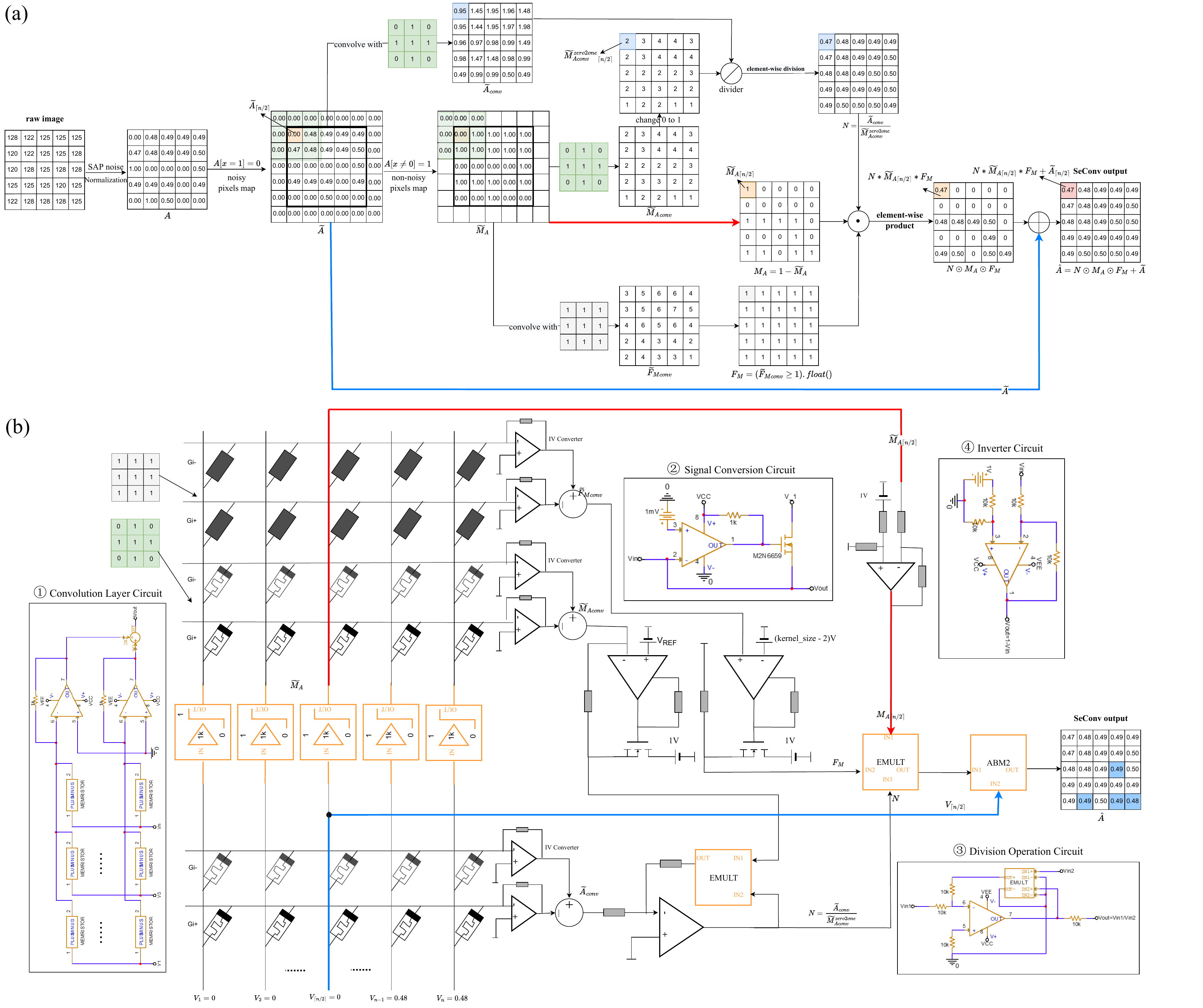}
  \caption{Schematic of the SeConv model with a \(3\times 3\) convolution kernel and a \(5\times 5\) input tensor. (a) illustrates the denoising principle and process, where the preprocessed noisy image through the SeConv model yields the restored tensor. (b) shows the circuitry of the SeConv model, the input is kept consistent with (a), and the output results of both are compared, noting that the errors remain within an acceptable range. In (b), errors are denoted with blue markings.}
  \label{fig_2}
\end{figure*}

\begin{figure}
  \centering
  \includegraphics[width=3.4in]{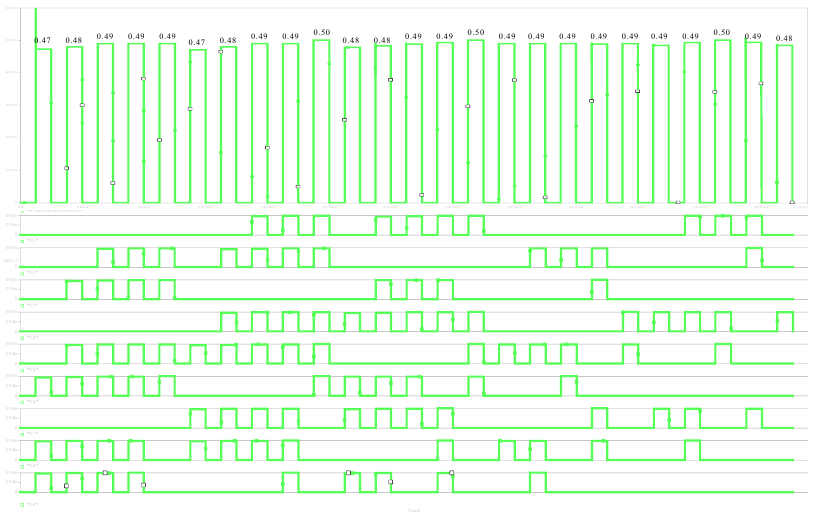}
  \caption{Simulation results of circuit in Fig. \ref{fig_2}(b).}
  \label{fig_30}
\end{figure}

The raw image is introduced into SAP noise to generate a noisy image and normalized to 0 to 1, resulting in tensor \(A\). The tensor \(A\) is then preprocessed into \(\widetilde{A}\), which serves as the input for the SeConv model. The theoretical noise removal results are shown in Fig. \ref{fig_2}(a), while PSpice circuit of the memristor-based SeConv (MSC) model and its simulation results are presented in Fig. \ref{fig_2}(b) and Fig. \ref{fig_30}, respectively. These results serve to validate the effectiveness and feasibility of the circuit. The weights used in Fig. \ref{fig_2}, both in theoretical calculations and circuit simulations, are obtained by ternarizing the weights of the \(3\times 3\) convolutional kernels from a pre-trained SeConvNet.


\subsection{The Processing Steps of SeConv Model}
The processing steps of the SeConv model with a \(3\times 3\) kernel on a noisy image are depicted in Fig. \ref{fig_2}(a). Convolution operations on the input tensor \(\widetilde{A}\) and the non-noisy pixel map \(\widetilde{M}_A\) generate tensors \(\widetilde{A}_{conv}\) and \(\widetilde{M}_{Aconv}\) respectively, while fixed convolution on the \(\widetilde{M}_A\) generate tensors \(\widetilde{F}_{Mconv}\). Subsequently, 0s in \(\widetilde{M}_{Aconv}\) are substituted with 1s, resulting in tensor \(\widetilde{M}_{Aconv}^{zero2one}\). This tensor is then used for element-wise division with \(\widetilde{A}_{conv}\), yielding tensor \(N\). Tensor \(M_A\) is obtained by subtracting \(\widetilde{M}_A\) from 1. The element greater than or equal to 1 in tensor \(\widetilde{F}_{Mconv}\) is 1, otherwise it is 0, and tensor \(F_M\) is obtained. The product of \(N\), \(M_A\), and \(F_M\) is added to the input tensor \(\widetilde{A}\) to derive the result of the SeConv model, which represents the recovered tensor.

\subsection{Convolution Layer Circuit}
To achieve the convolution as described in \eqref{eq3} and \eqref{eq4}, we employ a memristor-based convolution (MC) layer and resistor-based convolution (RC) layer circuit depicted in portions {\large \ding{172}} of Fig. \ref{fig_2}(b), where weights are represented by memristors or resistors.

The input voltage is transformed into currents through two columns of memristors or resistors (denoted as \(G_+\) and \(G_-\)), resulting in \(I_+\) and \(I_-\) after accumulation. Voltage signals \(V_+\) and \(V_-\) are then obtained through a current-to-voltage converter. Finally, the result of a convolution layer is derived through a subtractor, i.e., \(V_{out} = V_+ - V_-\).

\subsection{Signal Conversion Circuit}
The signal conversion circuit is depicted in portions {\large \ding{173}} of Fig. \ref{fig_2}(b). The two inputs of the comparator are connected to the reference and input signals respectively, while its output is connected to the gate of a metal-oxide-semiconductor field-effect transistor (MOSFET). The drain of the MOSFET is supplied with a 1V voltage, and the source is connected to the input signal.

When the reference signal value is lower than the input signal value, a low-level output is generated by the comparator, causing the MOSFET to turn off and setting the output of the circuit to be equivalent to the input. Conversely, a high-level output is produced by the comparator, causing the MOSFET to turn on and the circuit output to be 1V. Thus, by setting the reference signal to 0V or close to 0V, input signal voltage values of 0V can be converted to 1V while preserving other values.

To implement the algorithm described in \eqref{eq4}, the reference signal is connected to the negative input terminal of the comparator and set to \(\eta\)V, while the source of the MOSFET is grounded.

The input voltage \(V_{in}\) represents the element values in the tensor \(\widetilde{M}_{Aconv}\), may theoretically be negative but is still converted to 1V in the circuit, potentially introducing errors. However, we have chosen the design in portions {\large \ding{173}} of Fig. \ref{fig_2}(b) as errors are negligible, and ensuring positive voltage is crucial for subsequent circuits. Thus, this circuit design is considered essential.

\subsection{Division Operation Circuit}
The division operation circuit is composed of an operational amplifier and a multiplier, as depicted in portions {\large \ding{174}} of Fig. \ref{fig_2}(b). \(V_{in2}\) must be positive, while \(V_{in1}\) can be either positive or negative. \(V_{in2}\) is the output voltage of the signal conversion circuit, and the condition \(V_{in2} > 0\) always holds.

\subsection{Inverter Circuit}
The inverter circuit depicted in portions {\large \ding{175}} of Fig. \ref{fig_2}(b) was designed to convert the non-noisy pixels map \(\widetilde{M}_A\) into the noisy pixels map \(M_A\). \(\widetilde{M}_A\) is a binary tensor where 0s denote noisy pixel coordinates and 1s denote non-noisy pixel coordinates. The input signal is connected to the negative input terminal of the operational amplifier, while the positive input terminal is connected to a 1V voltage. Consequently, the circuit's output \(V_{out}\) is calculated as \(V_{out}=-V_{in}+1\).

\subsection{A Novel Memristive Circuit Architecture}
The evaluation of the sign of \(\widetilde{M}_{Aconv}\) in the circuit is difficult. Thus, we convert the elements in the tensor \({\widetilde{M}}_{Aconv}\) that are 0 to 1 as

\begin{equation}
  \label{eq13}
  \left[{\widetilde{M}}_{Aconv}^{zero2one}\right]_{i,j,k} =
  \begin{cases}
    1 & \text{if } \left[{\widetilde{M}}_{Aconv}\right]_{i,j,k}=0 \\
    \left[{\widetilde{M}}_{Aconv}\right]_{i,j,k} & \text{otherwise}.
  \end{cases}
\end{equation}

We can reformulate \eqref{eq3} as

\begin{equation}
  \label{eq14}
  \left[N\right]_{i,j,k}=\frac{\left[{\widetilde{A}}_{conv}\right]_{i,j,k}}{\left[{\widetilde{M}}_{Aconv}^{zero2one}\right]_{i,j,k}}.
\end{equation}

The signal conversion circuit was designed specifically to retain the analog values of non-zero elements in the tensor \({\widetilde{M}}_{Aconv}\) while only converting the 0s. In addition, to maximize the restoration of noisy pixels, the values of the tensor \(F_M\) are set to 1. We can reformulate \eqref{eq6} as

\begin{equation}
  \label{eq15}
  \hat{A}=\widetilde{A}+N\odot M_A.
\end{equation}

\begin{figure}[!t]
  \centering
  \includegraphics[width=3.4in]{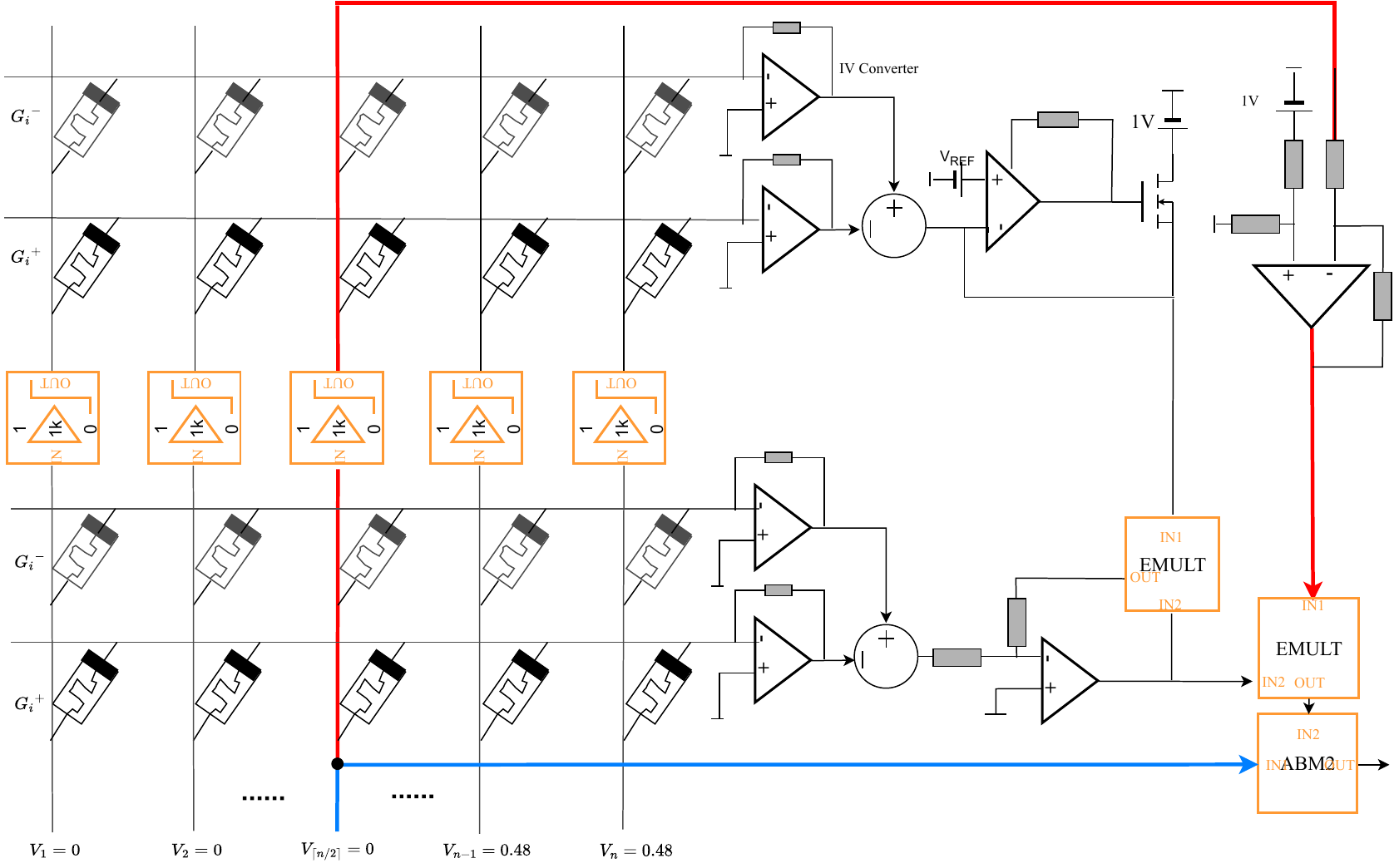}
  \caption{The circuit architecture of the MSCE model. The red and blue lines highlight the portions of the circuit that only select the central voltage of \(\widetilde{M}_A\) and \(\widetilde{A}\).}
  \label{fig_3}
\end{figure}

The circuit of the enhanced MSC (MSCE) model is depicted in Fig. \ref{fig_3}. Specifically, the convolution result \({\widetilde{A}}_{conv}\) is generated by the input signal \(\widetilde{A}\) through MC. Then, \({\widetilde{M}}_A\) is produced by \(\widetilde{A}\) passing through a gain limiter and then convolved through MC to obtain \({\widetilde{M}}_{Aconv}\). Through the signal conversion circuit, \({\widetilde{M}}_{Aconv}^{zero2one}\) is generated. At the divider, where the value of \(N\) is calculated according to \eqref{eq14}. Simultaneously, the central voltage signal \({\widetilde{M}}_{A\left\lceil n/2\right\rceil}\) is used to generate a noisy pixel map \(M_{A\left\lceil n/2\right\rceil}\) through the inverter circuit, where \(n\) represents the quantity of input signals and \(\left\lceil\cdot\right\rceil\) is the ceiling function. Finally, at the multiplier, \(M_{A\left\lceil n/2\right\rceil}\) and \(N\) are combined, and their output, along with the voltage signal \(V_{\left\lceil n/2\right\rceil}\) from \(\widetilde{A}\), converge at the adder to form the output of circuit. The circuit operates in parallel.

Note that the portions highlighted by the red and blue lines in Fig. \ref{fig_2} and Fig. \ref{fig_3}. Single signals (\(M_{A\left\lceil n/2\right\rceil}\) and \(V_{\left\lceil n/2\right\rceil}\)) are exclusively chosen for computation. The reason is that the circuit performs a single convolution operation, resulting in a single value for \(\widetilde{A}_{conv}\) and \(\widetilde{M}_{Aconv}\). \(\widetilde{M}_A\) can be obtained without accumulation operations. Consequently, the dimensions of \(\widetilde{A}_{conv}\) and \(\widetilde{M}_{Aconv}\) do not match \(\widetilde{M}_A\), thus precluding element-wise product. However, it is observed that the result of a single convolution operation corresponds to the element at the central position of the feature map region. Although multiple elements of \(\widetilde{M}_A\) are obtained simultaneously, only the one at the central position contributes to subsequent computations. Similarly, in the adder involving input signals, only the value at the central position is considered. In addition, given that the kernel size \(s\) is invariably odd, there consistently exists a central value that satisfies these criteria.

\section{EXPERIMENTAL ANALYSIS}
\label{s4}
\subsection{Overview}
In this study, denoising experiments were conducted on grayscale images. Firstly, the pre-trained full precision SeConv (FPSC) was ternarized according to \eqref{eq7}, resulting in the ternary SeConv (TSC) model. Then, ternary weights were mapped to the MSC model. In test-time, \(100\times 100\) images were randomly cropped from a part of the Berkeley segmentation dataset (BSD68) \cite{martin2001database}, and SAP noise was introduced, followed by preprocessing. In brief, after normalizing to 0 to 1, all values of 1 were set to 0 to ensure uniform noise values. Moreover, we compared the MSCE with the MSC in circuits to validate its effectiveness.

Simulation efficiency is significantly enhanced by co-simulation with MATLAB and PSpice. Image reading is managed by MATLAB, and image processing is conducted by the PSpice circuit model in the Simulink toolbox. The processed data is subsequently saved and converted to grayscale for further analysis in subsequent comparative experiments.


\begin{figure*}
  \centering
  \includegraphics[width=7.0in]{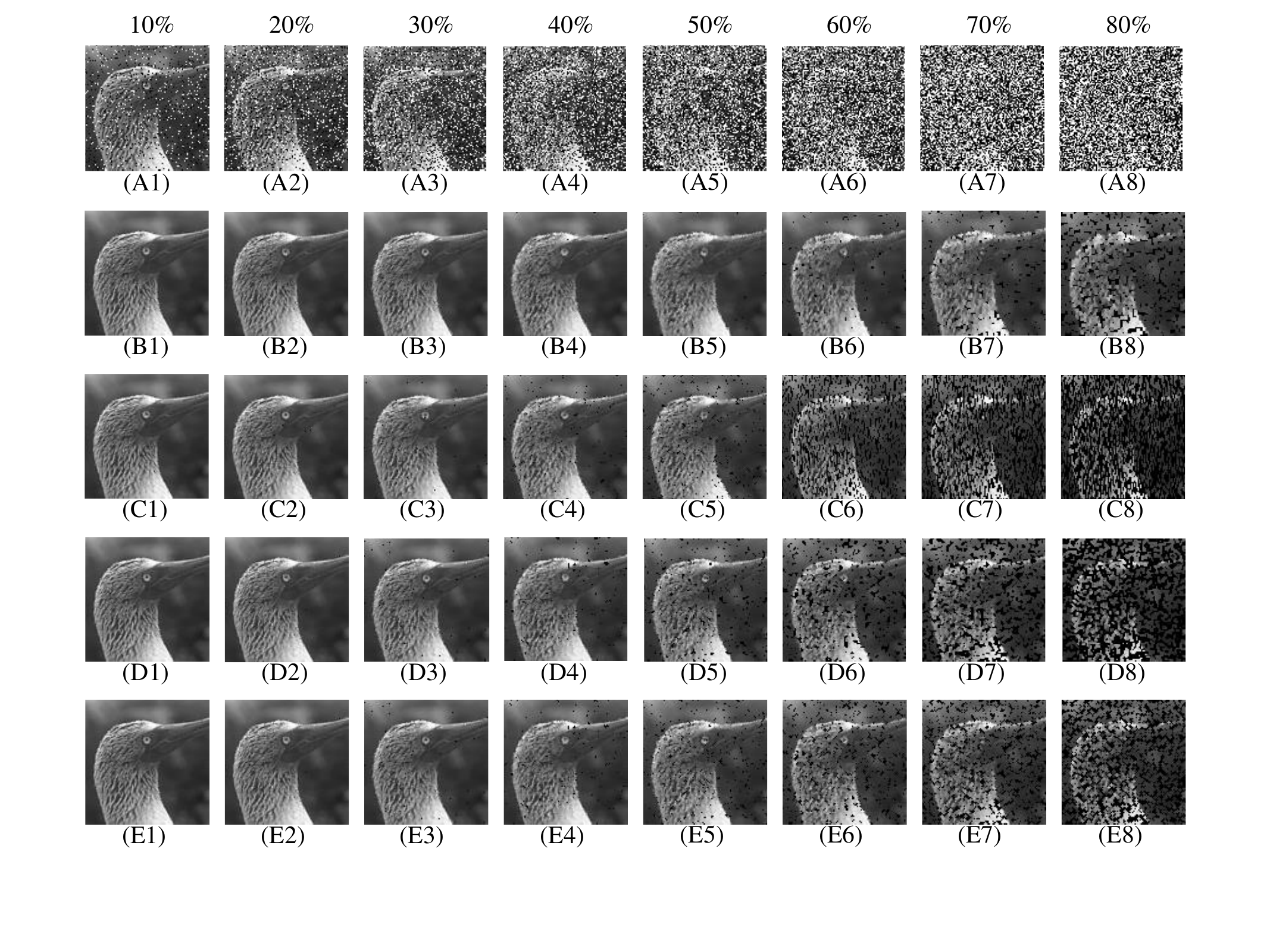}
  \caption{Restoration results of different models. A1-A8 depict noisy images, while B1-B8, C1-C8, D1-D8, and E1-E8 represent restoration images by the FPSC, TSC, MSC, and MSCE models respectively.}
  \label{fig_4}
\end{figure*}

\begin{table*}[!t]
  \centering
  \caption{Results of PSNR (\textnormal{d}B) and SSIM for different models at different noise densities}
  \label{table_1}
  \renewcommand{\arraystretch}{1.25}
  \setlength{\tabcolsep}{1.6\tabcolsep}
  \begin{tabular}{|c|c|c|c|c|c|c|c|c|c|c|}
    \hline
    Criteria & Models & 10\% & 20\% & 30\% & 40\% & 50\% & 60\% & 70\% & 80\% & Mean \\
    \hline
    \multirow{4}{*}{\rotatebox{90}{PSNR}} & FPSC & 33.78 & 30.24 & \textbf{28.39} & \textbf{26.23} & \textbf{24.99} & \textbf{22.19} & \textbf{18.44} & \textbf{14.97} & \textbf{24.90} \\
    \cline{2-11}
    & TSC & \textbf{34.35} & 30.47 & 27.17 & 24.36 & 21.21 & 14.22 & 12.23 & 10.60 & 21.83 \\
    \cline{2-11}
    & MSC & 34.05 & 30.49 & 27.11 & 23.51 & 20.31 & 16.82 & 13.44 & 10.71 & 22.06 \\
    \cline{2-11}
    & MSCE & 34.17 & \textbf{30.53} & 27.25 & 23.77 & 21.08 & 17.89 & 14.71 & 12.03 & 22.68 \\
    \hline
    \multirow{4}{*}{\rotatebox{90}{SSIM}} & FPSC & 0.982 & \textbf{0.962} & \textbf{0.939} & \textbf{0.905} & \textbf{0.854} & \textbf{0.715} & \textbf{0.528} & \textbf{0.274} & \textbf{0.77} \\
    \cline{2-11}
    & TSC & \textbf{0.985} & 0.958 & 0.909 & 0.805 & 0.645 & 0.242 & 0.156 & 0.105 & 0.601 \\
    \cline{2-11}
    & MSC & 0.982 & 0.961 & 0.9 & 0.795 & 0.598 & 0.391 & 0.223 & 0.114 & 0.621 \\
    \cline{2-11}
    & MSCE & 0.984 & \textbf{0.962} & 0.903 & 0.802 & 0.622 & 0.430 & 0.270 & 0.151 & 0.641 \\
    \hline
  \end{tabular}
\end{table*}

\begin{figure}[!t]
  \centering
  \includegraphics[width=3.4in]{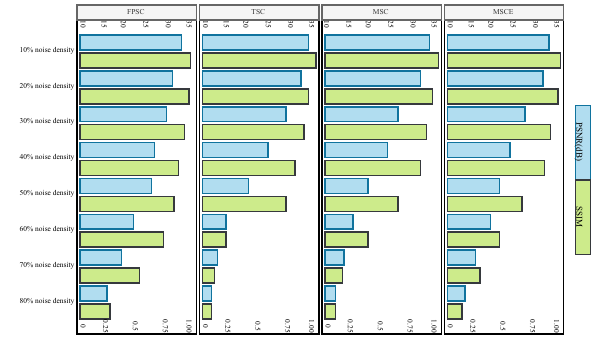}
  \caption{Comparison of the different models at different noise densities.}
  \label{fig_5}
\end{figure}

\begin{figure}[!t]
  \centering
  \includegraphics[width=3.4in]{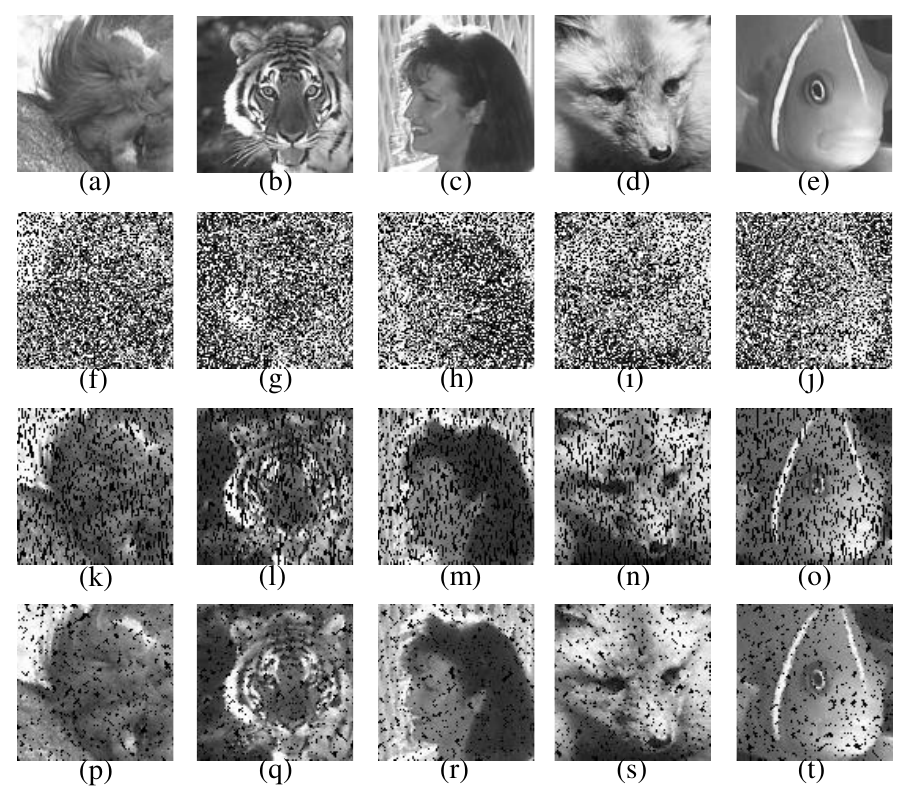}
  \caption{Restoration results of the TSC and MSCE models at 60\% noise density. a-e depict raw images, f-j depict noisy images, while k-o and p-t represent restored images by the TSC and MSCE models respectively.}
  \label{fig_6}
\end{figure}

\begin{figure}
  \centering
  \includegraphics[width=3.4in]{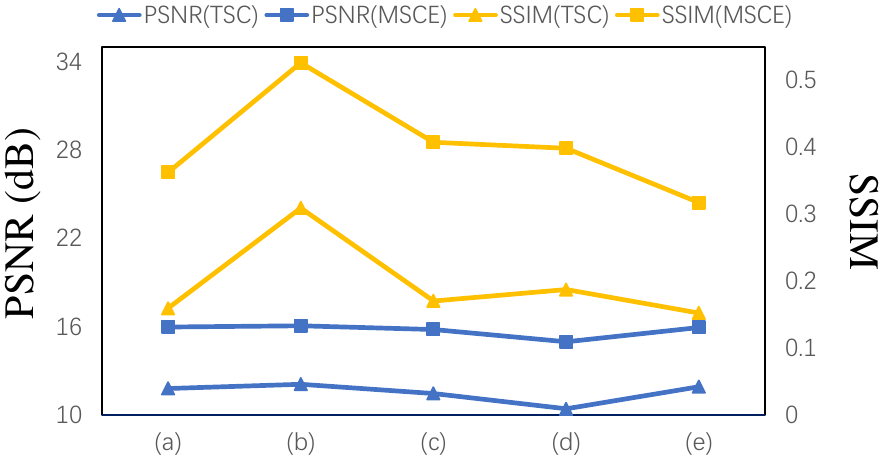}
  \caption{Results of the TSC and MSCE models at 60\% noise density.}
  \label{fig_50}
\end{figure}

\begin{table}[!t]
  \centering
  \caption{Results of PSNR (\textnormal{d}B) and SSIM for the TSC and MSCE models at 60\% noise density}
  \label{table_2}
  \renewcommand{\arraystretch}{1.25}
  \setlength{\tabcolsep}{0.95\tabcolsep}
  \begin{tabular}{|c|c|c|c|c|c|c|c|}
    \hline
    Criteria & Models & (a) & (b) & (c) & (d) & (e) & Mean \\
    \hline
    \multirow{2}{*}{\rotatebox{90}{PSNR}} & TSC & 11.83 & 12.12 & 11.49 & 10.45 & 11.95 & 11.57 \\
    \cline{2-8}
    & MSCE & \textbf{16.01} & \textbf{16.08} & \textbf{15.84} & \textbf{15.00} & \textbf{15.97} & \textbf{15.78} \\
    \hline
    \multirow{2}{*}{\rotatebox{90}{SSIM}} & TSC & 0.160 & 0.310 & 0.171 & 0.188 & 0.153 & 0.196 \\
    \cline{2-8}
    & MSCE & \textbf{0.363} & \textbf{0.527} & \textbf{0.408} & \textbf{0.399} & \textbf{0.318} & \textbf{0.403} \\
    \hline
  \end{tabular}
\end{table}

\subsection{Experiments on SAP Noise Removal}
In this section, the SAP noise removal performance of the FPSC, TSC, MSC, and MSCE models is evaluated across noise densities ranging from 10\% to 80\% using a randomly selected image from the BSD68 dataset, as illustrated in Fig. \ref{fig_4}. It is observed that good performance is maintained by the FPSC model at 60\% noise density, while the TSC, MSC, and MSCE models demonstrate satisfactory performance even at 50\% noise density.

The SAP noise removal performance of the FPSC, TSC, MSC, and MSCE models is presented in Table \ref{table_1}. It is observed that the denoising efficacy of the TSC model is slightly lower compared to the FPSC model because of the ternary weights. The TSC model serves as a benchmark for performance comparison of the MSC and MSCE models. Note that both the MSC and MSCE models outperform the TSC model when noise density is greater than or equal to 60\%, with the MSCE model slightly outperforming the MSC model. This trend is depicted in Fig. \ref{fig_5}. Further discussion on the enhancement of both MSC and MSCE models will be provided in the subsequent section.

\subsection{The Exploration of Circuit Performance Enhancement}
To verify the sustained superiority of the MSCE model over the TSC model at high noise density (60\% or greater), grayscale images were randomly selected from the BSD68 dataset, and tested at 60\%, 70\%, and 80\% noise densities. Consistent advantages of the MSCE model over the TSC model were observed at these high noise densities. For instance, Fig. \ref{fig_6} depicts the restoration results of both models at 60\% noise density. The PSNR (dB) and SSIM are presented in Table \ref{table_2} and Fig. \ref{fig_50}.

\begin{figure}[!t]
  \centering
  \includegraphics[width=3.4in]{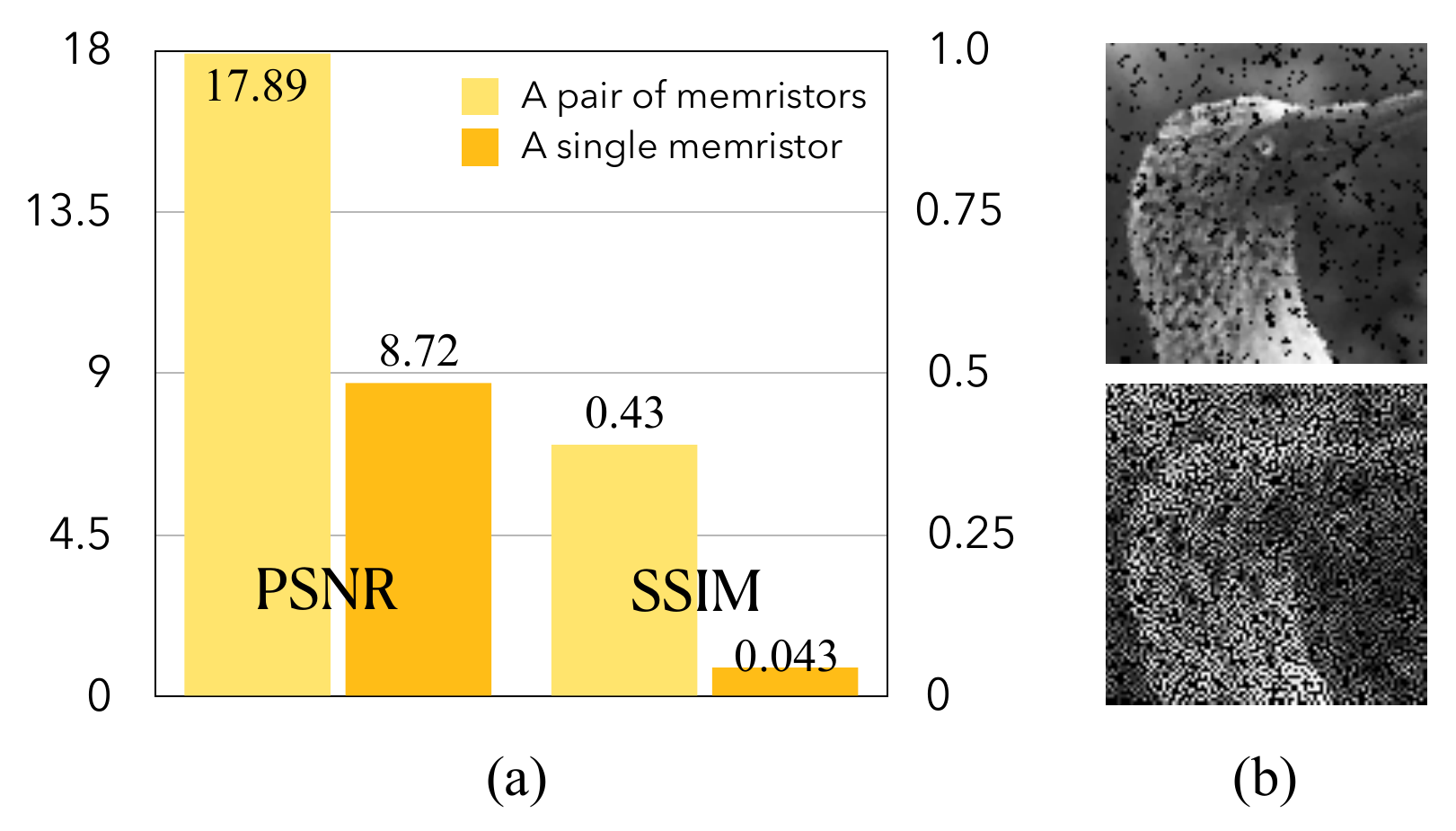}
  \caption{Restoration results of different structures at 60\% noise density, (a) shows PSNR and SSIM when ternary weights are represented by a pair of memristors and a single memristor respectively, (b) displays the restored images for both structures.}
  \label{fig_7}
\end{figure}

The differential structure in the MSCE circuit is believed to further reduce noise on a theoretical basis, enhancing the robustness of MSCE at high noise density. To validate this hypothesis, the ternary weights from a pair of memristors were replaced by a single memristor. Testing was conducted using kernels containing only non-negative values, with \(G_{on}\) representing 1 and \(G_{off}\) representing 0. The results of the test at 60\% noise density are depicted in Fig. \ref{fig_7}. The model performance is significantly improved under the same conditions by the differential structure. Therefore, the differential structure in the circuit is considered to be one of the reasons for the superior performance of MSCE.

\subsection{Power Consumption Analysis}
In the MSC and the MSCE models used in this study, there are eight pairs of memristors with a weight value of 0 and ten pairs of memristors with a weight value of 1. The power consumption of the memristor can be obtained by

\begin{equation}
  \label{eq16}
  P=\frac{W}{t}=\frac{\int_{t_0}^{t_1}{{v_{\left(t\right)}}_1^2R_{\left(t\right)}^{-1}}d_t+\int_{t_1}^{t_2}{{v_{\left(t\right)}}_2^2R^{-1}}d_t}{t_2-t_0}.
\end{equation}

The initial resistance of all memristors is set to \(R_{ON}\). When the programming voltage is 2V, the programming power consumption of a single memristor from \(R_{ON}\) to \(R_{OFF}\) is 15.7\(\mu\)W. As a result, the total power consumed during programming for all memristors in the MSC and MSCE models will not surpass 565.2\(\mu\)W.

The input voltage to the circuit, while it is running, is consistent with the value of tensor \(\widetilde{A}\), that is, 0 to 1V. The power consumption at various input voltages is shown in Tables \ref{table_3}, corresponding to \({\widetilde{W}}_i\) values of 0 and 1/-1. For pairs of memristors or resistors with a single input voltage representing a weight of 0 or 1, the power consumption in the MSC model (two pairs of memristors and a pair of resistors) is 103.36 and 234.09 \(\mu\)W, respectively, and 2.36 and 133.09 \(\mu\)W in the MSCE model (two pairs of memristors), respectively. 

For the \(3\times 3\) kernel with nine input voltages in the MSCE model as shown in Fig. \ref{fig_3}, the power consumption is 674.9\(\mu\)W, while the MSC model in Fig. \ref{fig_2}, it amounts to 1583.9\(\mu\)W. On average, each input voltage consumes about 74.98 and 175.98\(\mu\)W in the MSCE and MSC models, respectively, reducing to 0 when the input voltage is 0. Thus, the power consumption of all memristors and resistors can be estimated when processing \(100 \times 100\) images at different noise densities, as shown in Table \ref{table_4}. The power consumption of the MSCE model reduces by 57.6\% compared with the MSC model, and it has better denoising performance.

\begin{table*}[!t]
  \centering
  \caption{The power consumption(\(\mu\)W) of memristors at different input voltages}
  \label{table_3}
  \renewcommand{\arraystretch}{1.25}
  \setlength{\tabcolsep}{1.25\tabcolsep}
  \begin{tabular}{|c|c|c|c|c|c|c|c|c|c|c|c|}
    \hline
    \({\widetilde{W}}_i\) & Models & 0.1V & 0.2V & 0.3V & 0.4V & 0.5V & 0.6V & 0.7V & 0.8V & 0.9V & Mean \\
    \hline
    \multirow{2}{*}{0} 
    & MSC & 103.02 & 103.08 & 103.18 & 103.32 & 103.5 & 103.72 & 103.98 & 104.28 & 104.62 & 103.36 \\
    \cline{2-12}
    & MSCE & \textbf{2.02} & \textbf{2.08} & \textbf{2.18} & \textbf{2.32} & \textbf{2.5} & \textbf{2.72} & \textbf{2.98} & \textbf{3.28} & \textbf{3.62} & \textbf{2.36} \\
    \hline
    \multirow{2}{*}{1/-1}
    & MSC & 204.01 & 206.04 & 211.09 & 218.16 & 227.25 & 238.36 & 251.49 & 266.64 & 283.81 & 234.09 \\
    \cline{2-12}
    & MSCE & \textbf{103.01} & \textbf{105.04} & \textbf{110.09} & \textbf{117.16} & \textbf{126.25} & \textbf{137.36} & \textbf{150.49} & \textbf{165.64} & \textbf{182.81} & \textbf{133.09} \\
    \hline
  \end{tabular}
\end{table*}

\begin{table}[!t]
  \centering
  \caption{The power consumption(W) for processing \(100\times 100\) images at different noise densities}
  \label{table_4}
  \renewcommand{\arraystretch}{1.25}
  \setlength{\tabcolsep}{0.7\tabcolsep}
  \begin{tabular}{|c|c|c|c|c|c|c|c|c|}
    \hline
    Models & 10\% & 20\% & 30\% & 40\% & 50\% & 60\% & 70\% & 80\% \\
    \hline
    MSC & 1.58 & 1.41 & 1.23 & 1.06 & 0.88 & 0.70 & 0.53 & 0.35 \\
    \hline
    MSCE & \textbf{0.67} & \textbf{0.60} & \textbf{0.52} & \textbf{0.45} & \textbf{0.37} & \textbf{0.30} & \textbf{0.22} & \textbf{0.15} \\
    \hline
  \end{tabular}
\end{table}

\section{CONCLUSION}
\label{s5}
In this study, a novel memristor-based SAP noise removal circuit was proposed. Compared with existing memristive circuits, the MSC model can effectively restore SAP noise-corrupted images at high noise density. The MSCE model, obtained by improving the MSC model, exhibits superior performance, with power consumption reducing by 57.6\% compared with the latter. In addition, the MSCE model performs better than its theoretical value, the TSC model, at high noise density. Specifically, the PSNR/SSIM values between the MSCE model and the TSC model differ by no more than 0.59dB/0.023 at noise densities of up to 50\%. However, at noise densities of 60\%, 70\%, and 80\%, the PSNR/SSIM values of the MSCE model surpass those of the TSC model by 3.67dB/0.188, 2.48dB/0.114, and 1.43dB/0.046, respectively. A similar phenomenon is observed in the MSC model. The enhanced robustness and performance of the MSC and MSCE models at high noise density are attributed to the differential structure in the memristive circuit. Further optimization and improvement of related algorithms and circuit structures will be pursued in subsequent research to address higher-density noise. We believe that this study provides a useful prospect for SAP noise removal circuit design at high noise density.

\bibliographystyle{IEEEtran}
\bibliography{references}

\begin{thebibliography}{10}
\providecommand{\url}[1]{#1}
\csname url@samestyle\endcsname
\providecommand{\newblock}{\relax}
\providecommand{\bibinfo}[2]{#2}
\providecommand{\BIBentrySTDinterwordspacing}{\spaceskip=0pt\relax}
\providecommand{\BIBentryALTinterwordstretchfactor}{4}
\providecommand{\BIBentryALTinterwordspacing}{\spaceskip=\fontdimen2\font plus
\BIBentryALTinterwordstretchfactor\fontdimen3\font minus \fontdimen4\font\relax}
\providecommand{\BIBforeignlanguage}[2]{{%
\expandafter\ifx\csname l@#1\endcsname\relax
\typeout{** WARNING: IEEEtran.bst: No hyphenation pattern has been}%
\typeout{** loaded for the language `#1'. Using the pattern for}%
\typeout{** the default language instead.}%
\else
\language=\csname l@#1\endcsname
\fi
#2}}
\providecommand{\BIBdecl}{\relax}
\BIBdecl

\bibitem{xu2017fast}
S.~Xu, X.~Yang, and S.~Jiang, ``A fast nonlocally centralized sparse representation algorithm for image denoising,'' \emph{Signal Processing}, vol. 131, pp. 99--112, 2017.

\bibitem{ghimpecteanu2015decomposition}
G.~Ghimpe{\c{t}}eanu, T.~Batard, M.~Bertalm{\'\i}o, and S.~Levine, ``A decomposition framework for image denoising algorithms,'' \emph{IEEE transactions on Image Processing}, vol.~25, no.~1, pp. 388--399, 2015.

\bibitem{li2013computation}
P.~Li, D.~J. Lilja, W.~Qian, K.~Bazargan, and M.~D. Riedel, ``Computation on stochastic bit streams digital image processing case studies,'' \emph{IEEE Transactions on Very Large Scale Integration (VLSI) Systems}, vol.~22, no.~3, pp. 449--462, 2013.

\bibitem{khateb20150}
F.~Khateb, T.~Kulej, and M.~Kumngern, ``0.5-v dtmos median filter,'' \emph{AEU-International Journal of Electronics and Communications}, vol.~69, no.~11, pp. 1733--1736, 2015.

\bibitem{zhang2018modified}
S.~Zhang, X.~Li, and C.~Zhang, ``Modified adaptive median filtering,'' in \emph{2018 International Conference on Intelligent Transportation, Big Data \& Smart City (ICITBS)}.\hskip 1em plus 0.5em minus 0.4em\relax IEEE, 2018, pp. 262--265.

\bibitem{hwang1995adaptive}
H.~Hwang and R.~A. Haddad, ``Adaptive median filters: new algorithms and results,'' \emph{IEEE Transactions on image processing}, vol.~4, no.~4, pp. 499--502, 1995.

\bibitem{meher2014improved}
S.~K. Meher and B.~Singhawat, ``An improved recursive and adaptive median filter for high density impulse noise,'' \emph{AEU-International Journal of Electronics and Communications}, vol.~68, no.~12, pp. 1173--1179, 2014.

\bibitem{erkan2018different}
U.~Erkan, L.~G{\"o}krem, and S.~Engino{\u{g}}lu, ``Different applied median filter in salt and pepper noise,'' \emph{Computers \& Electrical Engineering}, vol.~70, pp. 789--798, 2018.

\bibitem{monajati2015approximate}
M.~Monajati, S.~M. Fakhraie, and E.~Kabir, ``Approximate arithmetic for low-power image median filtering,'' \emph{Circuits, Systems, and Signal Processing}, vol.~34, pp. 3191--3219, 2015.

\bibitem{memics2021different}
S.~Memi{\c{s}} and U.~Erkan, ``Different adaptive modified riesz mean filter for high-density salt-and-pepper noise removal in grayscale images,'' \emph{Avrupa Bilim ve Teknoloji Dergisi}, no.~23, pp. 359--367, 2021.

\bibitem{zhang2014new}
P.~Zhang and F.~Li, ``A new adaptive weighted mean filter for removing salt-and-pepper noise,'' \emph{IEEE signal processing letters}, vol.~21, no.~10, pp. 1280--1283, 2014.

\bibitem{thanh2020adaptive}
D.~N. Thanh, N.~N. Hien, P.~Kalavathi, and V.~S. Prasath, ``Adaptive switching weight mean filter for salt and pepper image denoising,'' \emph{Procedia Computer Science}, vol. 171, pp. 292--301, 2020.

\bibitem{erkan2020improved}
U.~Erkan, D.~N. Thanh, S.~Engino{\u{g}}lu, and S.~Memi{\c{s}}, ``Improved adaptive weighted mean filter for salt-and-pepper noise removal,'' in \emph{2020 International Conference on Electrical, Communication, and Computer Engineering (ICECCE)}.\hskip 1em plus 0.5em minus 0.4em\relax IEEE, 2020, pp. 1--5.

\bibitem{enginouglu2020adaptive}
S.~Engino{\u{g}}lu, U.~Erkan, and S.~Memi{\c{s}}, ``Adaptive ces{\'a}ro mean filter for salt-and-pepper noise removal,'' \emph{El-Cezeri}, vol.~7, no.~1, pp. 304--314, 2020.

\bibitem{varatharajan2018adaptive}
R.~Varatharajan, K.~Vasanth, M.~Gunasekaran, M.~Priyan, and X.~Z. Gao, ``An adaptive decision based kriging interpolation algorithm for the removal of high density salt and pepper noise in images,'' \emph{Computers \& Electrical Engineering}, vol.~70, pp. 447--461, 2018.

\bibitem{khateb2016low}
F.~Khateb, M.~Kumngern, S.~B.~A. Dabbous, and T.~Kulej, ``Low-voltage low-power bulk-driven analog median filter,'' \emph{AEU-International Journal of Electronics and Communications}, vol.~70, no.~5, pp. 698--706, 2016.

\bibitem{monajati2019modified}
M.~Monajati and E.~Kabir, ``A modified inexact arithmetic median filter for removing salt-and-pepper noise from gray-level images,'' \emph{IEEE Transactions on Circuits and Systems II: Express Briefs}, vol.~67, no.~4, pp. 750--754, 2019.

\bibitem{vasicek2016evolutionary}
Z.~Vasicek, V.~Mrazek, and L.~Sekanina, ``Evolutionary functional approximation of circuits implemented into fpgas,'' in \emph{2016 IEEE Symposium Series on Computational Intelligence (SSCI)}.\hskip 1em plus 0.5em minus 0.4em\relax IEEE, 2016, pp. 1--8.

\bibitem{yildirim2021analog}
M.~Yildirim, ``Analog circuit implementation based on median filter for salt and pepper noise reduction in image,'' \emph{Analog Integrated Circuits and Signal Processing}, vol. 107, no.~1, pp. 195--202, 2021.

\bibitem{zhang2018residual}
Y.~Zhang, Y.~Tian, Y.~Kong, B.~Zhong, and Y.~Fu, ``Residual dense network for image super-resolution,'' in \emph{Proceedings of the IEEE conference on computer vision and pattern recognition}, 2018, pp. 2472--2481.

\bibitem{zhang2021plug}
K.~Zhang, Y.~Li, W.~Zuo, L.~Zhang, L.~Van~Gool, and R.~Timofte, ``Plug-and-play image restoration with deep denoiser prior,'' \emph{IEEE Transactions on Pattern Analysis and Machine Intelligence}, vol.~44, no.~10, pp. 6360--6376, 2021.

\bibitem{valsesia2020deep}
D.~Valsesia, G.~Fracastoro, and E.~Magli, ``Deep graph-convolutional image denoising,'' \emph{IEEE Transactions on Image Processing}, vol.~29, pp. 8226--8237, 2020.

\bibitem{zhang2018ffdnet}
K.~Zhang, W.~Zuo, and L.~Zhang, ``Ffdnet: Toward a fast and flexible solution for cnn-based image denoising,'' \emph{IEEE Transactions on Image Processing}, vol.~27, no.~9, pp. 4608--4622, 2018.

\bibitem{zhang2017beyond}
K.~Zhang, W.~Zuo, Y.~Chen, D.~Meng, and L.~Zhang, ``Beyond a gaussian denoiser: Residual learning of deep cnn for image denoising,'' \emph{IEEE transactions on image processing}, vol.~26, no.~7, pp. 3142--3155, 2017.

\bibitem{zhang2015salt}
X.~Zhang, F.~Ding, Z.~Tang, and C.~Yu, ``Salt and pepper noise removal with image inpainting,'' \emph{AEU-International Journal of Electronics and Communications}, vol.~69, no.~1, pp. 307--313, 2015.

\bibitem{xing2019deep}
Y.~Xing, J.~Xu, J.~Tan, D.~Li, and W.~Zha, ``Deep cnn for removal of salt and pepper noise,'' \emph{IET Image Processing}, vol.~13, no.~9, pp. 1550--1560, 2019.

\bibitem{rafiee2023very}
A.~A. Rafiee and M.~Farhang, ``A very fast and efficient multistage selective convolution filter for removal of salt and pepper noise,'' \emph{Journal of Ambient Intelligence and Humanized Computing}, vol.~14, no.~9, pp. 1--17, 2023.

\bibitem{rafiee2023deep}
------, ``A deep convolutional neural network for salt-and-pepper noise removal using selective convolutional blocks,'' \emph{Applied Soft Computing}, vol. 145, p. 110535, 2023.

\bibitem{suresh2019realizing}
B.~Suresh, P.~K.~R. Boppidi, B.~P. Rao, S.~Banerjee, and S.~Kundu, ``Realizing spike-timing dependent plasticity learning rule in pt/cu: Zno/nb: Sto memristors for implementing single spike based denoising autoencoder,'' \emph{Journal of Micromechanics and Microengineering}, vol.~29, no.~8, p. 085006, 2019.

\bibitem{zarandi2020memristor}
A.~D. Zarandi, A.~Rubio, and M.~R. Reshadinezhad, ``A memristor-based quaternary memory with adaptive noise tolerance,'' in \emph{2020 XXXV Conference on Design of Circuits and Integrated Systems (DCIS)}.\hskip 1em plus 0.5em minus 0.4em\relax IEEE, 2020, pp. 1--6.

\bibitem{zhu2020convolution}
S.~Zhu, L.~Wang, Z.~Dong, and S.~Duan, ``Convolution kernel operations on a two-dimensional spin memristor cross array,'' \emph{Sensors}, vol.~20, no.~21, p. 6229, 2020.

\bibitem{shang2018srmc}
L.~Shang, S.~Duan, L.~Wang, and T.~Huang, ``Srmc: A multibit memristor crossbar for self-renewing image mask,'' \emph{IEEE Transactions on Very Large Scale Integration (VLSI) Systems}, vol.~26, no.~12, pp. 2830--2841, 2018.

\bibitem{duan2024memristor}
X.~Duan, Z.~Cao, K.~Gao, W.~Yan, S.~Sun, G.~Zhou, Z.~Wu, F.~Ren, and B.~Sun, ``Memristor-based neuromorphic chips,'' \emph{Advanced materials}, p. 2310704, 2024.

\bibitem{kataeva2015efficient}
I.~Kataeva, F.~Merrikh-Bayat, E.~Zamanidoost, and D.~Strukov, ``Efficient training algorithms for neural networks based on memristive crossbar circuits,'' in \emph{2015 International Joint Conference on Neural Networks (IJCNN)}.\hskip 1em plus 0.5em minus 0.4em\relax IEEE, 2015, pp. 1--8.

\bibitem{krestinskaya2019neuromemristive}
O.~Krestinskaya, A.~P. James, and L.~O. Chua, ``Neuromemristive circuits for edge computing: A review,'' \emph{IEEE transactions on neural networks and learning systems}, vol.~31, no.~1, pp. 4--23, 2019.

\bibitem{alibart2013pattern}
F.~Alibart, E.~Zamanidoost, and D.~B. Strukov, ``Pattern classification by memristive crossbar circuits using ex situ and in situ training,'' \emph{Nature communications}, vol.~4, no.~1, p. 2072, 2013.

\bibitem{hasan2014enabling}
R.~Hasan and T.~M. Taha, ``Enabling back propagation training of memristor crossbar neuromorphic processors,'' in \emph{2014 International Joint Conference on Neural Networks (IJCNN)}.\hskip 1em plus 0.5em minus 0.4em\relax IEEE, 2014, pp. 21--28.

\bibitem{soudry2015memristor}
D.~Soudry, D.~Di~Castro, A.~Gal, A.~Kolodny, and S.~Kvatinsky, ``Memristor-based multilayer neural networks with online gradient descent training,'' \emph{IEEE transactions on neural networks and learning systems}, vol.~26, no.~10, pp. 2408--2421, 2015.

\bibitem{zhang2023edge}
W.~Zhang, P.~Yao, B.~Gao, Q.~Liu, D.~Wu, Q.~Zhang, Y.~Li, Q.~Qin, J.~Li, Z.~Zhu \emph{et~al.}, ``Edge learning using a fully integrated neuro-inspired memristor chip,'' \emph{Science}, vol. 381, no. 6663, pp. 1205--1211, 2023.

\bibitem{radlak2020deep}
K.~Radlak, L.~Malinski, and B.~Smolka, ``Deep learning based switching filter for impulsive noise removal in color images,'' \emph{Sensors}, vol.~20, no.~10, p. 2782, 2020.

\bibitem{thanh2019iterative}
D.~N.~H. Thanh, S.~Eng{\'\i}no{\u{g}}lu \emph{et~al.}, ``An iterative mean filter for image denoising,'' \emph{IEEE Access}, vol.~7, pp. 167\,847--167\,859, 2019.

\bibitem{aguirre2024hardware}
F.~Aguirre, A.~Sebastian, M.~Le~Gallo, W.~Song, T.~Wang, J.~J. Yang, W.~Lu, M.-F. Chang, D.~Ielmini, Y.~Yang \emph{et~al.}, ``Hardware implementation of memristor-based artificial neural networks,'' \emph{Nature Communications}, vol.~15, no.~1, p. 1974, 2024.

\bibitem{li2016ternary}
F.~Li, B.~Liu, X.~Wang, B.~Zhang, and J.~Yan, ``Ternary weight networks,'' \emph{arXiv preprint arXiv:1605.04711}, 2016.

\bibitem{pershin2012spice}
Y.~V. Pershin and M.~Di~Ventra, ``Spice model of memristive devices with threshold,'' \emph{arXiv preprint arXiv:1204.2600}, 2012.

\bibitem{ran2020shufflenetv2}
H.~Ran, S.~Wen, S.~Wang, Y.~Cao, P.~Zhou, and T.~Huang, ``Memristor-based edge computing of shufflenetv2 for image classification,'' \emph{IEEE Transactions on Computer-Aided Design of Integrated Circuits and Systems}, vol.~40, no.~8, pp. 1701--1710, 2020.

\bibitem{ran2020memristor}
H.~Ran, S.~Wen, Q.~Li, Y.~Yang, K.~Shi, Y.~Feng, P.~Zhou, and T.~Huang, ``Memristor-based edge computing of blaze block for image recognition,'' \emph{IEEE Transactions on Neural Networks and Learning Systems}, vol.~33, no.~5, pp. 2121--2131, 2020.

\bibitem{chen2021highly}
J.~Chen, S.~Wen, K.~Shi, and Y.~Yang, ``Highly parallelized memristive binary neural network,'' \emph{Neural Networks}, vol. 144, pp. 565--572, 2021.

\bibitem{wen2018memristive}
S.~Wen, H.~Wei, Z.~Zeng, and T.~Huang, ``Memristive fully convolutional network: An accurate hardware image-segmentor in deep learning,'' \emph{IEEE Transactions on Emerging Topics in Computational Intelligence}, vol.~2, no.~5, pp. 324--334, 2018.

\bibitem{wen2019memristor}
S.~Wen, H.~Wei, Z.~Yan, Z.~Guo, Y.~Yang, T.~Huang, and Y.~Chen, ``Memristor-based design of sparse compact convolutional neural network,'' \emph{IEEE Transactions on Network Science and Engineering}, vol.~7, no.~3, pp. 1431--1440, 2019.

\bibitem{martin2001database}
D.~Martin, C.~Fowlkes, D.~Tal, and J.~Malik, ``A database of human segmented natural images and its application to evaluating segmentation algorithms and measuring ecological statistics,'' in \emph{Proceedings Eighth IEEE International Conference on Computer Vision. ICCV 2001}, vol.~2.\hskip 1em plus 0.5em minus 0.4em\relax IEEE, 2001, pp. 416--423.

\end{thebibliography}

\end{document}